\documentclass{article}
\usepackage{IEEEtrantools}
\usepackage{spconfa4,amsmath,graphicx}
\usepackage{booktabs}
\usepackage{cite}
\usepackage{subcaption}
\usepackage{amssymb}
\usepackage{color}
\usepackage{mathtools}
\usepackage{siunitx}
\usepackage{multirow}
\usepackage{makecell}

\title{EXPLOITING NOISE INSEPARABILITY FOR WEAKLY-SUPERVISED\\ DISCRIMINATIVE SPEECH DENOISING USING NOISY TARGETS}

\twoauthors{Matthew Maciejewski}{Johns Hopkins University\\ Human Language Technology Center of Excellence\\ Baltimore, MD, USA}{Samuele Cornell}{Carnegie Mellon University\\ Language Technologies Institute\\ Pittsburgh, PA, USA}

\usepackage[absolute,overlay]{textpos}
\newcommand{\copyrightstatement}{
    \begin{textblock}{1.0}(0.09, 0.89)
         \noindent
         \footnotesize
         \copyright 2026 IEEE. Personal use of this material is permitted. Permission from IEEE must be obtained for all other uses, in any current or future media, including reprinting/republishing this material for advertising or promotional purposes, creating new collective works, for resale or redistribution to servers or lists, or reuse of any copyrighted component of this work in other works.
    \end{textblock}
}
\begin{document}
\copyrightstatement
\bstctlcite{IEEEexample:BSTcontrol}
\ninept

\maketitle
\begin{abstract} 
Speech denoising is an often necessary step not only for human listening, but also for downstream processing by systems lacking robustness to noisy, real-world acoustic conditions. Unfortunately, denoising is a problem where conventional in-domain supervised training is not trivial, as the training targets cannot be annotated by humans: producing a clean version of a naturally-noisy speech recording is itself the task to solve. Supervised training is typically performed through the artificial addition of noise to clean speech recordings, which can only be sourced from controlled domains, a significant limitation due to the poor out-of-domain generalization of neural networks. An alternative is noisy target training (NyTT), which simply replaces the clean speech with in-domain noisy recordings, with the hope that learning to remove the artificial noise will extend to the natural. Though having shown promising results, NyTT's training objective is not minimized by clean speech estimates. We show that by estimating the artificial noise in addition to the naturally-noisy speech, the undesirable optimum can actually be exploited: the residual noise in the speech estimate can be canceled by the noise estimate via simple subtraction. Crucially, the optimum is fully compatible with conventional artificial mixtures, enabling joint training using both types of data with consistent optimization targets, opening the door to improved domain adaptability. The effectiveness of our approach is demonstrated through WHAM! and CHiME-3-based benchmarks.
\end{abstract}

\section{Introduction}
\label{sec:introduction}
Speech enhancement is the task of producing a clean, high-fidelity waveform of a speech signal from a recording where the collected speech has been degraded in some way, for example, due to additive interferences (such as background noise or other speech), the conditions of the recording (such as reverberation or poor microphone quality), or even consequences of the processing chain (such as codec artifacts or sample rate).
In many situations, these degradations are unavoidable, and thus speech enhancement systems are used to improve the quality of the recordings, both for human listeners and as front-ends for downstream speech processing systems whose performance may be degraded by the corrupted speech.

Modern speech enhancement techniques based on deep neural networks are typically trained in a supervised manner, requiring a substantial amount of noisy-to-clean paired speech data~\cite{wang2018supervised}.
As a consequence, one of the more salient aspects of the task is a dependence on synthetic data.
In contrast to many other speech technologies, like speech recognition or speaker identification, it is not possible for a human to annotate the ground truth target---to be able to produce the underlying clean speech waveform from a degraded recording would be to be able to solve the speech enhancement task itself.
Thus, supervised deep learning systems have traditionally been trained using synthetic data: high-quality clean speech recordings are artificially degraded in ways that approximate the real-world degradations, and this artificially degraded speech is used alongside the original clean signal for supervised learning.

Unfortunately, this approach is not without its flaws.
It might be difficult to replicate the degradation process, and furthermore there can be domain mismatches from both the clean speech and the artificial degradations to the naturally-degraded speech~\cite{chime7udase}.
This can be problematic due to the tendency of neural networks to have poor generalization outside of the data they were trained on.

As a result, there is a need for enhancement methods that can effectively leverage naturally-collected speech data for which no corresponding clean ground truth signal is available.
Among discriminative approaches, noisy target training~(NyTT)~\cite{nytt,nytt_analysis} is currently the most prominent paradigm. 
Drawing from the Noise2Noise framework originally proposed for image denoising~\cite{noise2noise}, in NyTT a network is trained to predict a noisy speech signal from a noisier version of the same signal, created by artificially adding more noise from the same domain. 
MixIT~\cite{mixit} and its extension RemixIT~\cite{remixit} also adopt the same noisy-target principle, with the latter additionally employing mean-teacher~\cite{mean_teacher} self-pseudo-labeling to improve performance.
A key practical advantage of these discriminative approaches is that they only need in-domain noise recordings, which are relatively easy to collect in most domains and applications.
Instead, generative weakly-supervised speech denoising methods~\cite{cyclegan_se,unse}, require access to an unpaired corpus of clean speech, which is difficult to obtain for many languages and spontaneous conversational speech.

As separating a mixture of two background noises is nearly impossible~\cite{noisy_oracle,ringmix}, we show that when estimating noisy speech in the NyTT framework, since the network is trained to output one noise (in the noisy speech target) but suppress another (the artificially-added one), the resulting network splits the difference, outputting both noises at a reduced amplitude along with the speech.
We note, however, that a similar effect happens if estimating the artificially-added noise, resulting in the estimate being the full noise mixture.
This noise estimate can then be subtracted off of the noisy-speech estimate to produce a fully-denoised speech estimate, which we leverage in an approach called \textit{Differential Noise Filtering}~(DNF).

This same two-output formulation is also compatible with fully-supervised training using synthetic mixtures, allowing joint training using weakly-supervised and fully-supervised data with consistent optimization targets, enabling systems that leverage both the power of simulated mixtures but also the domain-adaptability of NyTT.
We demonstrate the effectiveness of our approach on the WHAM! and CHiME-3 datasets, showing gains of up to \SI{5.9}{dB} over the NyTT baseline on WHAM!, and DNSMOS improvement of 1.01 on CHiME-3 real noisy speech compared to 0.44 of the best-performing baseline.

\section{Theory and Method}
\label{sec:theory_and_method}
The basic formulation of the speech denoising task is to produce an estimate $\hat{s}$ of a speech waveform $s \in \mathbb{R}^T$ from a noisy speech recording $x \in \mathbb{R}^T$, which can be modeled as a simple summation of $s$ with a noise signal $n \in \mathbb{R}^T$:
\begin{gather}
x = s + n \text{,} \label{eq:mix}
\end{gather}
where neither $s$ nor $n$ are observed directly. 

In traditional supervised learning, the samples used in training are generated artificially.
Training using naturally-collected mixtures $x$ would require having access to the underlying speech signal $s$ to use as a training target, but to generate that signal would mean already having the capability to perform speech denoising.
As a result, the typical approach is to collect clean (noiseless) speech $s$ and noise $n$ separately and sum them to form synthetic mixtures $x$.
However, as discussed in Section~\ref{sec:introduction}, while in-domain noise $n$ can be collected with any recording device, clean speech $s$ requires the much greater burden of a recording studio, which both restricts the amount of available training data and forces reliance on studio-quality speech, which is poorly representative of spontaneous conversational interactions, introducing substantial domain mismatch with real-world deployment conditions.

Following the NyTT~\cite{nytt} paradigm, we instead replace the clean speech in~(\ref{eq:mix}) with a naturally-noisy in-domain recording $s^{\text{noisy}}$:
\begin{equation}
x = s^\text{noisy} + n_2 = (s + n_1)+n_2 \text{,}\label{eq:noisy_mixture}
\end{equation}
where $n_2$ is artificially-added noise drawn from the same domain as the noise $n_1$ already present in $s^{\text{noisy}}$. The target remains $s$, even though only $s^{\text{noisy}}$ is available at training time. 
The benefit of this formulation, however, is that the training data can be collected from the target deployment domain.

\subsection{Signal Rescaling}
\label{ssec:signal_rescaling}

In this work, we rely heavily on the scale-invariance theory by Le~Roux et~al.~\cite{sisdr}.
Audio waveforms do not have a true ``scale'' associated with them, as the numerical values we assign to the shape resulting from the underlying physical pressure waves depend only on the digital processing chain.
As a result, when performing signal arithmetic for metrics and loss functions, it is important to ensure that the signal scale does not unintentionally impact the intended computation.

Le~Roux et~al. perform optimal rescaling by leveraging the fact that sampled audio waveforms can be approximated as zero-mean random processes, and so in expectation, the dot product of two unrelated waveforms is approximately zero, an approximation that holds well due to the tens of thousands of samples per second typical in audio recordings.
They then note that identical scaling can be achieved of two arbitrarily-scaled versions of a waveform even when one also contains an \textit{additional} unrelated signal.
More precisely, for uncorrelated $x$ and $y$ in $\mathbb{R}^T$, if observations $\alpha x$ and $\beta(x+y)$ are made with $\alpha$ and $\beta$ being arbitrary scaling coefficients,
\begin{equation}
\frac{||\alpha x||^2}{\langle\alpha x, \beta(x+y)\rangle} = \frac{\alpha^2 ||x||^2}{\alpha\beta ||x||^2} = \frac{\alpha}{\beta}\text{,}
\end{equation}
and so recovering the relative scaling of $\alpha$ and $\beta$ is possible, meaning the signals can be rescaled with identical scaling of $x$ in both.

We note, however, that correct scaling \textit{cannot} be achieved if a non-$x$ signal is present in \textit{both} observations (i.e.\ $\alpha(x+z)$ and $\beta(x+y)$), which both constrains the signal pairs we can operate on and dictates where estimation errors should live.
If estimation errors are present, they should be confined to the signal with the extra component (e.g.\ $x+y$), as they can be considered part of the extra component. Errors in the single-element signal (e.g.\ $x$) not only corrupt the scaling computation but can introduce degenerate solutions.

\subsection{Loss Analysis}\label{ssec:loss_analysis}

We follow a similar approach to Maciejewski \& Cornell~\cite{ringmix}, who analyze a similar formulation for noisy speech separation and find theoretically and experimentally that the Scale-Invariant Signal-to-Distortion Ratio~(SI-SDR)~\cite{sisdr} loss function is minimized by speech estimates that include the full mixture noise at half amplitude.
This follows from an assumption that networks can discriminate between speech and noise, but are incapable of separating mixtures of background noise that follow the same distribution.
Under this assumption, given the noisy mixture formulation of equation~\eqref{eq:noisy_mixture}, a reasonable set of possible noisy speech estimates $\tilde{s}^\text{noisy}$ would be:
\begin{gather}
\tilde{s}^\text{noisy} \in \mathcal{S} \coloneq \{s + \lambda_s(n_1+n_2)\ |\ \lambda_s \in [0, 1]\} \text{,}\label{eq:enh_set}
\end{gather}
meaning the network can identify and arbitrarily scale the noise mixture $n_1+n_2$ by some factor $\lambda_s$, but cannot adjust the relative amount of $n_1$ and $n_2$ in the estimate.

We evaluate this set under the scale-dependent signal-to-distortion ratio (SDR)~\cite{sisdr} loss:
\begin{equation}
\ell_\text{SDR}(\hat{s}; s) \coloneq -10 \log \frac{||s||^2}{||s-\hat{s}||^2} \text{,}\label{eq:sdr}
\end{equation}
and compute the optimal scale $\lambda_s^*$ of residual noise that minimizes this loss function:
\begin{align}
\lambda_s^* & = \arg\min_{\lambda_s} \ell_\text{SDR}(\tilde{s}^\text{noisy}; s^\text{noisy}) \\
& = \arg\min_{\lambda_s} ||s^\text{noisy}-\tilde{s}^\text{noisy}||^2 \\
& = \arg\min_{\lambda_s}||(1-\lambda_s)n_1 - \lambda_s n_2||^2 \text{.}
\end{align}
Assuming $\langle n_1, n_2 \rangle \approx 0$ following the theory in Section~\ref{ssec:signal_rescaling},
\begin{align}
\lambda_s^* & = \arg\min_{\lambda_s} \left[ (1-\lambda_s)^2||n_1||^2 + \lambda_s^2 ||n_2||^2 \right] \\
& = \frac{||n_1||^2}{||n_1||^2 + ||n_2||^2}\text{.}
\end{align}
Here we see that under the NyTT framework, the speech estimate is incentivized to also output residual noise, proportional to how much of the noise mixture comes from the noisy speech target.

We note however, that a similar analysis could be performed if an estimate $\hat{n}$ of $n_2$ is produced by the network.
Then, our set is:
\begin{gather}
\tilde{n} \in \mathcal{N} \coloneq \{\lambda_n(n_1+n_2)\ |\ \lambda_n \in [0, 1]\} \text{,}
\end{gather}
and deriving the optimal scaling $\lambda_n^*$ is similar to before, resulting in:
\begin{gather}
\lambda_n^* = \frac{||n_2||^2}{||n_1||^2 + ||n_2||^2}\text{.}
\end{gather}
And additionally, if $\lambda_s$ and $\lambda_n$ were constrained to be equivalent, the derivation would result in a more stable optimum at exactly $\lambda = 0.5$, as long as one noise is not considerably louder than the other~\cite{ringmix}.

\subsection{Proposed Method}
\label{ssec:proposed_method}

\begin{figure}[t]
\centering
\includegraphics[width=1.0\columnwidth]{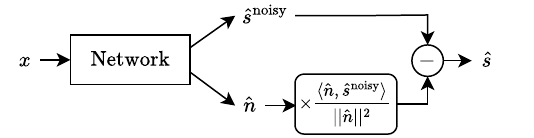}
\vspace{-16pt}
\caption{Proposed DNF framework, showing three estimated signals.}
\vspace{-10pt}
\label{fig:system}
\end{figure}

In this work, we propose to exploit this bias by training a two-output network that estimates both the noisy speech and the noise. 
When the NyTT optimization results in the same $\lambda(n_1+n_2)$ term in both outputs, their difference cancels exactly: $\tilde{s}^\text{noisy}-\tilde{n} = s$.
We refer to this approach as \textit{Differential Noise Filtering}~(DNF), in analogy with the differential amplifiers of electronic circuits.

As noted in the previous section, the stable $\lambda=0.5$ optimum only exists when both $(n_1+n_2)$ terms are constrained to have the same scaling.
So, we design a scaling strategy that aligns with the desired optimum.
Since $s$, $n_1$, and the estimation error should be orthogonal to $n_2$, we can rescale following the theory of Section~\ref{ssec:signal_rescaling} so that both estimates contain $0.5n_2$.
This results in the overall loss:
\begin{equation}
\ell_\text{noisy} = \ell_\text{SDR}\left( \frac{0.5 ||n_2||^2}{\langle n_2, \hat{s}^\text{noisy}\rangle}\hat{s}^\text{noisy}; s^\text{noisy} \right) + \ell_\text{SDR}\left(\frac{0.5||n_2||^2}{\langle n_2, \hat{n}\rangle}\hat{n}; n_2 \right)\hspace{-0.1em}\text{.}\label{eq:loss_noisy}
\end{equation}

However, this approach only works during training, where the supervision signals are known.
At inference time, the estimates can simply be scaled to each other, as there is no risk of learning degenerate solutions.
We scale the noise estimate to have the same amplitude as the noise in the noisy speech estimate, and $\hat{s}$ is thus:
\begin{gather}
\hat{s} \coloneq \hat{s}^\text{noisy} - \frac{\langle \hat{n}, \hat{s}^\text{noisy}\rangle}{||\hat{n}||^2}\hat{n} \text{.}\label{eq:s_est}
\end{gather}
This constitutes the proposed DNF architecture, shown in Figure~\ref{fig:system}.

Crucially, the proposed dual-branch framework can also be trained on conventional synthetic mixtures~\eqref{eq:mix}, allowing joint training on weakly-supervised and fully-synthetic data.
With synthetic data, the network can be \textit{directly} trained to produce estimates conducive to the DNF framework.
When constructing mixtures $x$ from $s$ and $n$, rather than training to estimate $s$ as usual, we use $s+0.5n$ and $0.5n$ as the supervision targets.
Additionally, the final DNF output $\hat{s}$ can be used directly with SI-SDR loss to improve estimate fidelity.
Overall, the loss for samples mixed using clean speech is:
\begin{equation}
\ell_\text{clean} = \ell_\text{SI-SDR}(\hat{s}^\text{noisy}; s+0.5n)+\ell_\text{SI-SDR}(\hat{n}; n) + \ell_\text{SI-SDR}(\hat{s}; s)\text{,}\label{eq:loss_clean}
\end{equation}
where $\hat{s}$ is computed as defined in equation~\eqref{eq:s_est}.

Note that the original NyTT framework cannot fully exploit mixed synthetic and noisy-target training to resolve this bias: the two target types impose conflicting optima on similar input mixtures.
Instead, with the proposed differential framework, we can avoid this by giving each supervision type targets with consistent minima within the same subtraction-based processing chain.

\section{Experimental Setup}
\label{sec:experimental_setup}
\subsection{Model and Training Configuration}
\label{ssec:model}

We use a 4-block TF-GridNet~\cite{tfgridnet1, tfgridnet2} with $L=4$, $D=24$, $I=4$, $J=4$, $H=192$, following the hyperparameter notation of~\cite{tfgridnet2}.
The STFT uses a \SI{32}{ms} Hann window with \SI{8}{ms} hop, matching the noisy/reverberant separation setup at \SI{16}{kHz} described in~\cite{tfgridnet2}.

For the baseline system, we use a 1-output model, trained with SI-SDR~\cite{sisdr} loss to either $s$ or $s^\text{noisy}$ depending on if clean- or noisy-target supervision is available (serving as a NyTT~\cite{nytt} baseline in the latter case).
For the proposed system, we use a 2-output model, matching Figure~\ref{fig:system}, trained with either $\ell_\text{clean}$~\eqref{eq:loss_clean} or $\ell_\text{noisy}$~\eqref{eq:loss_noisy} for clean- and noisy-target supervision respectively.

\subsection{Data}

Our core experiments use the \SI{16}{kHz} WHAM!~\cite{wham} single-speaker enhancement dataset.
To simulate the case where we can only collect noisy speech, we use the original WHAM! mixtures as the $s^\text{noisy}$ signal (with the underlying signals $s$ and $n_1$ unavailable in training/validation), and assign each mixture an \textit{additional} noise recording from WHAM! to serve as $n_2$.
We also evaluate using the unmodified WHAM! test set with only one noise source, leading to a mismatch with the double-noise training set, aligning with a realistic NyTT deployment scenario.

We additionally perform two modifications: 1. To investigate mixed-condition training, we randomly partition the dataset into two disjoint sets: one that uses the above NyTT setup and one that uses the normal clean-target formulation.
2. As the NyTT formulation's addition of noise to existing noise results in very low SNRs, we additionally explore scaling down the noise signals to evaluate method effectiveness using more reasonable, less noisy data.

To study synthetic-real generalization, we evaluate the real noisy speech of CHiME-3~\cite{chime3}, where WSJ0~\cite{wsj0} text was read in noisy environments like cafes or buses.
Our closely matched (but out-of-domain) simulations of this condition use background noises from the CHiME-3 environments added to the clean, studio WSJ0 speech from WHAM!
For in-domain NyTT, we add background noises from the same environments to already-noisy CHiME-3 speech.

\subsection{Metrics}

For experiments using WHAM!, where ground truth clean speech is available, we measure SI-SDR improvement (SI-SDRi).
For experiments using CHiME-3, where ground truth is \textit{not} available, we measure DNSMOS~\cite{dnsmos}, UTMOS~\cite{utmos}, and Word Error Rate~(WER) using the OWSM~v3.1~\cite{owsm} transcription model, all obtained through the VERSA toolkit~\cite{versa}.

\section{Results and Discussion}
\label{sec:results_and_discussion}
\begin{table}[t]
\caption{SI-SDRi~[dB] using models trained with noisy targets. The 0.707 scale halves the energy of the two noises to match the single-noise test set. The 0.282 scale is roughly the \SI{10}{dB} condition in \cite{nytt}.}
\vspace{-0.7em}
\centering
\sisetup{
    reset-text-series = false,
    text-series-to-math = true,
    mode=text,
    round-mode=places,
    round-precision=1,
    table-number-alignment=center}
\begin{tabular}{ccSS}
\toprule
{\multirowcell{2.9}{\textbf{Noise}\\ \textbf{Scale}}} & & \multicolumn{2}{c}{\textbf{WHAM! Test Set}} \\ \cmidrule{3-4}
 & \textbf{System} & {Original (Unscaled)} & {2-Noise (Scaled)} \\ \midrule
\multirow{2}{*}{1.0} & {Baseline} & 1.4660241323760324 & 3.4501420321293454 \\
 & {DNF} & 5.146841375828682 & 7.260933110586988 \\ \cmidrule{1-4}
\multirow{2}{*}{0.707} & {Baseline} & 2.366659055243867 & 3.990381830651003 \\
 & {DNF} & 8.32558962521508 & 9.050967367990873 \\ \cmidrule{1-4}
\multirow{2}{*}{0.282} & {Baseline} & 10.066914172478603 & 5.092901041155235 \\
 & {DNF} & 10.531134074419077 & 8.119237982044792 \\
\bottomrule
\end{tabular}
\vspace{-1.0em}
\label{tab:scaling_results}
\end{table}

\begin{table*}[t]
\caption{SI-SDRi~[db] on WHAM! when the training data is partitioned into clean- and noisy-target subsets. In cases such as 90/0, training was performed only on the 90\% of the dataset with clean targets, and the remaining noisy-target 10\% was not used.}
\vspace{-0.7em}
\centering
\sisetup{
    reset-text-series = false,
    text-series-to-math = true,
    mode=text,
    round-mode=places,
    round-precision=1,
    table-number-alignment=center}
\begin{tabular}{cSSSSSSSSSSSS}
\toprule
 & \multicolumn{10}{c}{\textbf{\% of WHAM! Train Set Allocated to Clean-Target / Noisy-Target}} \\ \cmidrule{2-13}
 & {All Clean} & \multicolumn{2}{c}{90\% Clean} & \multicolumn{2}{c}{75\% Clean} & \multicolumn{2}{c}{50\% Clean} & \multicolumn{2}{c}{25\% Clean} & \multicolumn{2}{c}{10\% Clean}& {All Noisy} \\ \cmidrule(lr){2-2} \cmidrule(lr){3-4} \cmidrule(lr){5-6} \cmidrule(lr){7-8} \cmidrule(lr){9-10} \cmidrule(lr){11-12} \cmidrule(lr){13-13}
 & {100/0} & {90/10} & {90/0} & {75/25} & {75/0} & {50/50} & {50/0} & {25/75} & {25/0} & {10/90} & {10/0} & {0/100}\\ \midrule
Baseline & 15.162380093490977 & 15.20458614199046 & 14.945259579892852 & 14.961801436351573 & 14.976056327906088 & 11.811015262752383 & 14.85360540190925 & 3.417672990260743 & 14.366798221038616 & 1.8901397041014474 & 13.784998380270437 & 1.4660241323760324 \\
DNF & {--} & 15.237717866109646 & {--} & 15.179633715080058 & {--} & 15.162063409414724 & {--} & 14.454394565191702 & {--} & 13.35733221857432 & {--} & 5.146841375828682 \\
\bottomrule
\end{tabular}
\vspace{-1.0em}
\label{tab:full_results}
\end{table*}

\begin{table}[t]
\caption{Evaluation on CHiME-3 real noisy speech using models trained on synthetic mixtures of WSJ0 clean speech with CHiME-3 noise, and noisy-target mixtures of CHiME-3 real noisy speech with additional noise. Batches of size 28 were constructed using a fixed number of samples from each set, with 28/0 and 0/28 using only the synthetic and only the real data respectively.}
\vspace{-0.7em}
\centering
\sisetup{
    reset-text-series = false,
    text-series-to-math = true,
    mode=text,
    round-mode=places,
    round-precision=2,
    table-number-alignment=center}
\addtolength{\tabcolsep}{-0.07em}
\begin{tabular}{ccSSS[table-format=2.1,round-precision=1]}
\toprule
{\multirowcell{2.9}{\textbf{Synth. / Real}\\ \textbf{Per Batch}}} & & \multicolumn{3}{c}{\textbf{Test Metric}} \\ \cmidrule{3-5}
 & \textbf{System} & {DNSMOS$\uparrow$} & {UTMOS$\uparrow$} & {WER\%$\downarrow$} \\ \midrule
{--} & {None} & 1.4471069832260441 & 1.5624990406490507 & 25.416200565901164 \\\cmidrule{1-5}
\multirow{2}{*}{28/0} & {Baseline} & 1.8948099053370329 & 1.8870489038430251 & 75.6754622622886 \\
 & {DNF} & 2.0345118412779284 & 1.9432956002620392 & 76.54405474764756 \\ \cmidrule{1-5}
\multirow{2}{*}{24/4} & {Baseline} & 1.734346136417711 & 1.6623159438371657 & \bfseries 23.971836546686845 \\
 & {DNF} & \bfseries 2.45604759030778 & \bfseries 2.402813232886843 & 28.22530762650523 \\ \cmidrule{1-5}
\multirow{2}{*}{14/14} & {Baseline} & 1.6911894079874126 & 1.5935242624122858 & 24.187010594196224 \\
 & {DNF} & 2.105569215702176 & 1.9364132458900476 & 27.29157070474436 \\ \cmidrule{1-5}
\multirow{2}{*}{0/28} & {Baseline} & 1.589456579175669 & 1.5778575138612227 & 25.008883332236625 \\
 & {DNF} & 1.789495515360957 & 1.6197509816456668 & 31.565440547476477 \\
\bottomrule
\end{tabular}
\vspace{-1.0em}
\label{tab:chime_results}
\end{table}

Our base noisy-target results are in Table~\ref{tab:scaling_results}, where we train and test using additional noise added to WHAM! noisy mixtures, with both noises scaled to achieve varying SNRs, as well as evaluating on the original WHAM! condition.
Scaling 1.0 is the realistic deployment condition, where the resulting NyTT mixtures have twice as much noise as the source domain.
Scale 0.707 halves the energy to compensate for this, and scale 0.282 is a higher-SNR condition where NyTT is known to work (approximately the \SI{10}{dB} condition from \cite{nytt}).
In all cases, the proposed DNF method outperforms the baseline, with a maximum benefit of \SI{5.9}{dB} improvement on WHAM!
We note, however, that even with fairly high-SNR noisy-target training, the performance is still well below the \SI{15.2}{dB} of conventional clean-target supervised training (Table~\ref{tab:full_results}).

Table~\ref{tab:full_results} contains the first of our results investigating mixed-condition training (i.e. joint clean- and noisy-target training).
In these experiments we partition the WHAM! training and validation data into a portion using the original clean-target supervision and a portion using the noisy-target configuration.
Using all-synthetic data lets us investigate performance more precisely, but lacks the practical problem we aim to solve: generalizing models trained on simulated data to noisy-target data from a \textit{different} domain.

We again note that even the better-performing DNF system only achieves \SI{5.1}{dB} on the all-noisy-target data compared to the \SI{15.2}{dB} of all-clean-target data system.
However, we can see that DNF is very receptive to mixed-condition training---it can tolerate half of the dataset being noisy-target before showing degradation, and even with just 10\% clean-target data degrades by only \SI{1.8}{dB}.
In contrast, the baseline is not so tolerant, dropping by \SI{3.4}{dB} at half-and-half, and by \SI{13.3}{dB} with 10\% clean-target supervision.

Disappointingly, the DNF system generally performs similarly to the baseline system trained on only the clean-target data (discarding the noisy-target), but does outperform it at high clean-target supervision rates.
However, if mixed condition training is \textit{necessary}, i.e. when the goal is to adapt a system to a target domain where clean speech cannot be collected, the proposed DNF system shows clear improvements.
This is demonstrated using our experiments on the real noisy speech from CHiME-3, presented in Table~\ref{tab:chime_results}.
Here we combine simulated data (adding CHiME-3 noise to WSJ0 read speech) with noisy-target training using CHiME-3 real noisy speech.

Training on only the synthetic mixtures (28/0) yields some improvement in DNSMOS and UTMOS, but causes a sharp rise in Word Error Rate under the OWSM speech recognition model, indicating that the enhanced output contains processing artifacts that impact downstream recognition.
All models trained using noisy targets do not show this behavior, demonstrating the value of in-domain training.
And, although training on only noisy target data (0/28) does show signal improvement, it performs worse than the synthetic-only condition, confirming that learning denoising from only noisy data is of limited effectiveness.

As expected, we observe that the proposed DNF method brings the largest improvements in the mixed-condition setting.
In contrast, adding some noisy-target training data to the baseline results in a degradation across all metrics, likely due to inconsistent optimization targets.
The best numbers are when each batch consists mostly of simulated mixtures (24, compared to only 4 noisy-target).
In this scheme, the DNF system achieves 2.46 DNSMOS and 2.40 UTMOS, compared to only 1.89 on both metrics from the best-performing baseline.
Notably, that best-performing baseline is the synthetic-only (28/0) configuration, the same one whose outputs cause significant WER increase.

Unfortunately, all DNF systems show slight degradation on WER compared to the unprocessed audio (while baselines show slight improvement).
Part of this gap can be explained by an implicit \emph{observation adding}~\cite{iwamoto2022bad} effect in the baseline: because the NyTT speech estimate retains roughly half the mixture noise, its outputs more closely resemble the noisy mixtures OWSM was trained on, keeping them within the transcription model's training distribution.
DNF, by contrast, produces cleaner speech but also, as a tradeoff, more pronounced artifacts, and thus sits further from OWSM's multi-condition training distribution, resulting in increased WER.

\section{Conclusion}
\label{sec:conclusion}
Training speech denoising models using noisy speech has been a goal of the field, as it enables training models using data collected from the intended deployment domain, leading to the development of Noisy-Target Training~(NyTT).
We have demonstrated that the theoretic optimum of NyTT is inconsistent with conventionally-supervised training and develop Differential Noise Filtering~(DNF) to exploit this optimum, which not only results in improved noisy-target training, but calibrates targets with conventional supervision.
This allows mixed training that leverages the potent supervision of simulated mixtures with the domain-adaptability of noisy targets, which we have demonstrated on real noisy speech from CHiME-3.

\bibliographystyle{IEEEtran}
\bibliography{refs}

\end{document}